\documentclass[oneside,a4paper]{article}

\newcommand{\topk}{\textit{Top-k}\xspace}
\newcommand{\skyline}{\textit{Skyline}\xspace}
\newcommand{\fskyline}{\textit{Flexible-skyline}\xspace}
\newcommand{\fs}{\textit{F-skyline}\xspace}
\newcommand{\fdominance}{\textit{\(\mathcal{F}\)-dominance}\xspace}
\newcommand{\nd}{\textsc{nd}\xspace}
\newcommand{\po}{\textsc{po}\xspace}

\usepackage{pgfplots}
\usepackage{float}
\usepackage{xspace}
\usepackage{array}
\usepackage[hidelinks]{hyperref}

\usepackage{authblk}
\providecommand{\keywords}[1]{\textbf{\textit{Keywords:}} #1}

\title{A bird's eye view on Multi-Objective Optimization techniques in Relational Databases}
\author{Giuseppe Tortorelli}
\affil{Politecnico di Milano\\
Milan, Italy\\
\href{mailto:giuseppe.tortorelli@mail.polimi.it}{giuseppe.tortorelli@mail.polimi.it} }
\date{}

\begin{document}
\maketitle
\begin{abstract}
Multi-objective optimization is the problem of optimizing simultaneously multiple objective functions and several techniques exist to deal with this problem.
This paper aims to present the main methods that can be used to solve this issue in the context of relational databases. 
In particular, this work examines \topk query to get the \textit{k} best result from a dataset and \skyline query that provides a more general overview of the best results. 
We also discuss \fskyline, a new method designed to improve upon the previous techniques, mitigating their shortcomings.  
For each method, we describe the main characteristics and present an overview of the algorithms implementing such thecniques, while comparing advantages and disadvantages.
\end{abstract}
    

\section{Introduction}
\begin{figure}[H]
    \centering
    \begin{minipage}{.5\textwidth}
        \centering
        \begin{tabular}{|c | c | c |} 
            \hline
            Name & Experience (years) & Score \\ [0.5ex] 
            \hline
            JS & 3  & 9.8\\ 
            \hline
            FS & 2 & 7.8\\
            \hline
            PT & 6 & 7.3 \\
            \hline
            MMM & 5 & 6.2\\
            \hline
            NF & 9 & 5.7 \\
            \hline
            SS & 10 & 3 \\
            \hline
            MR & 9 & 2 \\
            \hline
            DR & 8 & 4.5 \\
            \hline
            ... & ... & ... \\ [1ex] 
            \hline
           \end{tabular}
           \caption{example photographer dataset}
           \label{fig:table}
    \end{minipage}%
    \begin{minipage}{.5\textwidth}
      \centering
      \includegraphics[width=10cm]{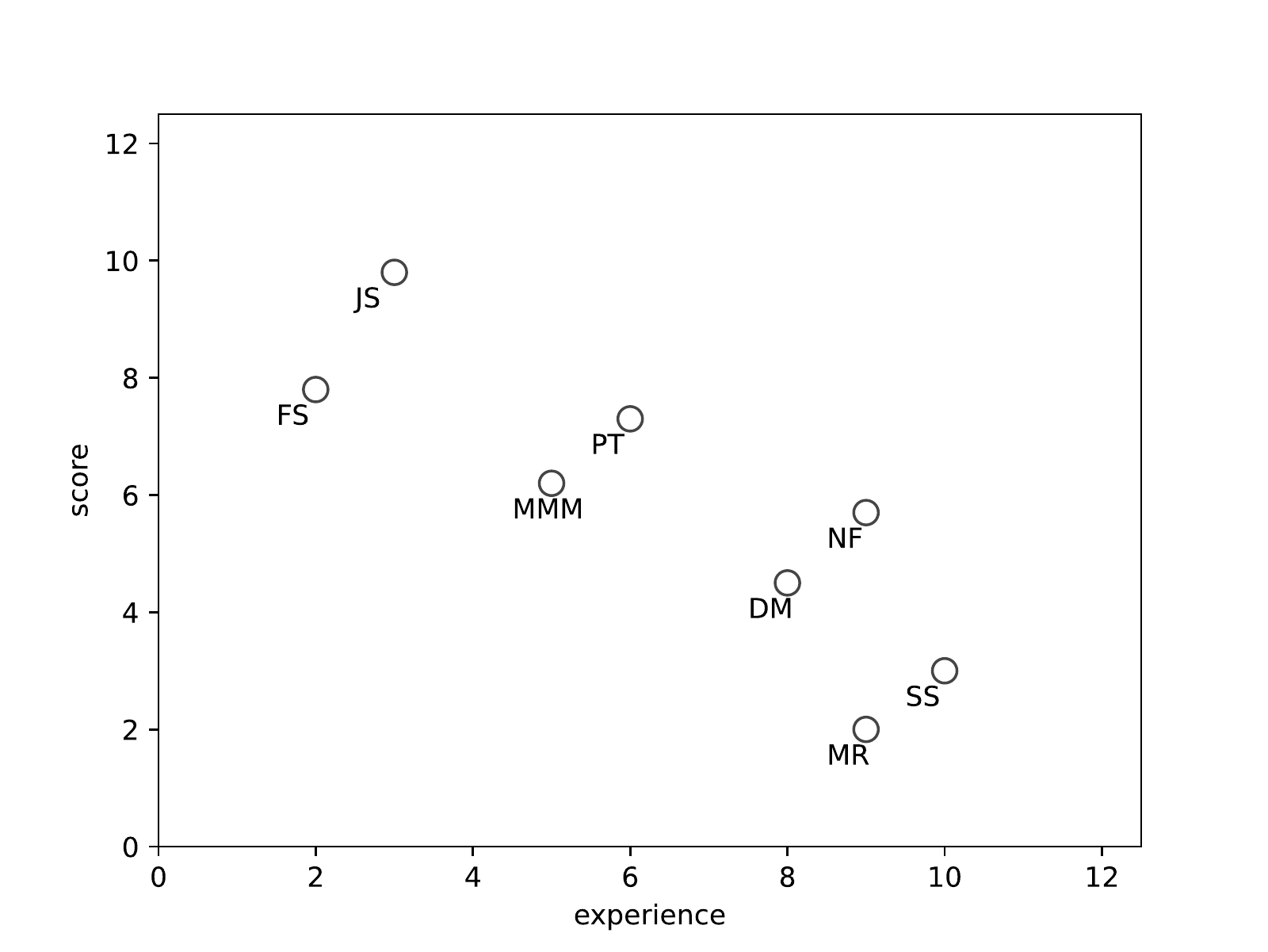}
      \caption{dataset geometric view}
      \label{fig:dataspace}
    \end{minipage}
\end{figure}

Being in the digital age means that data are becoming a fundamental part of everyday life since they harbor essential and valuable information. 
In modern applications, such as IoT \cite{Singh2021, 10.1007/s00779-012-0542-1}, Data Mining, and Machine Learning \cite{10.1145/1046456.1046467}, data comes in large quantities and in an unstructured way \cite{4812412}; dealing with this data highlights a considerable issue: how to transform such data into usable information by extracting interesting results from it.
The problem of retrieving the best results by combining different criteria is known as \textit{multi-objective optimization}.\hfill \break
To solve such a problem, different techniques have been proposed and the two most important paradigms are \topk and \skyline queries \cite{10.1145/1046456.1046467}.
The goal of \topk is to extract the best \textit{k} results by letting the user choose which attribute to optimize, whereas \skyline provides a general overview of the most relevant objects in the dataset. 
However, both have different pros and cons.\hfill \break
The goal of this paper is to square up a critical review of these approaches and to compare them to a new method: \fskyline.
This work is organized as follows: Section \hyperref[sec:topk]{2} contains explanations of \topk, Section \hyperref[sec:skyline]{3} discusses \skyline while Section \hyperref[sec:fskyline]{4} is about \fskyline.
Finally, in Section \hyperref[sec:comparison]{5} a comparison between them is presented.\hfill \break
The approach proposed in this paper provides not only a comparison but also a practical overview of these operators.
In this regard, each section contains a part in which general details  are provided  about the most used algorithms and a brief comparison between them.\hfill \break
To keep a practical approach, let us introduce an example that will be extended for \topk, \skyline, and \fskyline in the respective sections:\hfill \break
\break
\label{sec:ex1}\textit{Example 1}: Consider a company interested in finding
 a photographer for  advertising  its new product. 
The company consults a website that collects all freelancer video makers. 
In particular, the parameters they consider are the experience and the 
score obtained from the reviews.\hfill \break
Figure \hyperref[fig:table]{1} displays the freelancer dataset while each 
tuple is represented in a 2-dimensional geometric view in Figure 
\hyperref[fig:dataspace]{2}.\hfill \break

\section{Top-k}
\label{sec:topk}
\topk queries (\textit{ranking}) are one of the most used methods in multi-objective optimization \cite{10.1145/1391729.1391730} thanks to their efficiency and the ability to control the output size.\hfill \break
The core concept behind this approach is to transform the original multi-objective problem into a single-objective one, through the use of a scoring function, which aggregates the attribute of interest of the dataset.
In this way, we can select the \textit{k} results with the best score.\hfill \break
\break
\begin{figure}[h]
    \includegraphics[width=10cm]{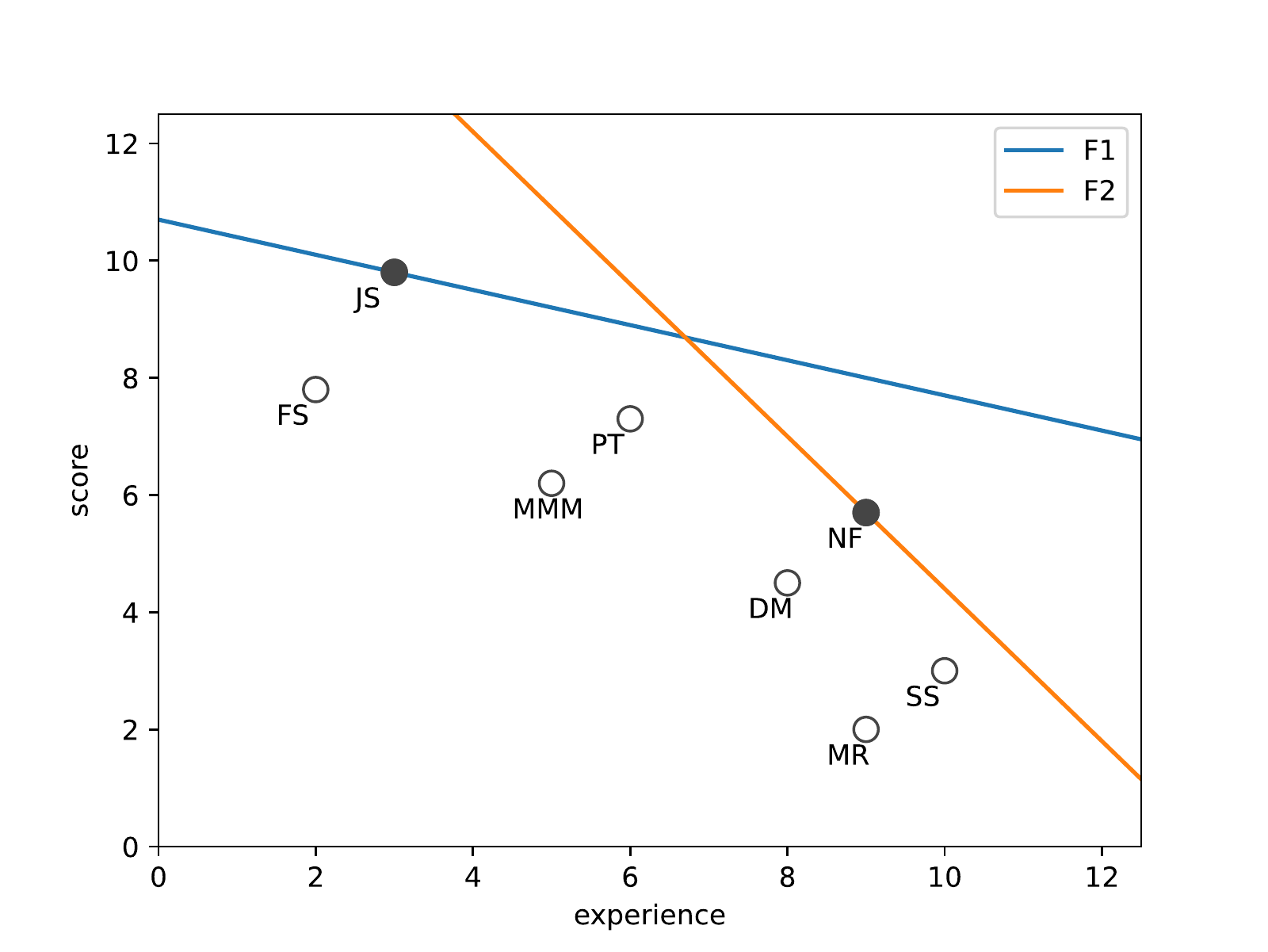}
    \centering
    \caption{\topk result view}
    \label{fig:extopk}
\end{figure}
\break
\textit{Example} \hyperref[sec:ex1]{1} \textit{(cont.)}. In order to select the best photographer, we consider two scoring functions as the weighted sum of the \textit{experience} and the \textit{score}:
\[ F1 = 0.6 · \ experience + 2 · \ score \] 
\[ F2 = 1.3 · \ experience + score \]
According to the selected weights, \textit{F1} gives more relevance to the \textit{score} while \textit{F2} prefers the \textit{experience}.
We want to find the most experienced photographer with the best score, which in mathematical terms means maximizing the functions.
Figure \hyperref[fig:extopk]{3} shows how the scoring function is represented in the geometrical view and their optimal values: JS is optimal for \textit{F1}, while NF is optimal for 
\textit{F2}.\hfill \break
\break
As shown in the previous example, the result set strongly depends on how the user selects the weights, whose choice is generally non-trivial.
For instance, if in \textit{F2} the experience had been favored a little more with respect to the score, the best result would have been SS instead of NF.\hfill \break

\subsection{Algorithms}
To talk about \topk algorithms for aggregations, we need to make 
a distinction between how data are accessed. 
We talk about \textit{sorted access} when we access objects sequentially in a list that has been previously ordered according to one or more attributes, \textit{random access} if objects are accessed directly in one step.\hfill \break
All of the following algorithms work for \textit{monotone} scoring functions. 
This is a reasonable property for a scoring function since
if the partial score \(\sigma[a^t_i]\)  of each attribute \textit{a} of a tuple $t$ is greater or equal to that of the attribute of a tuple $s$ \(\sigma[a^s_i]\), one would expect the overall score of $t$ \(\sigma[t]\)
to be greater or equal to the overall score of s \(\sigma[s]\).
$$
\sigma[a^t_i] \geq \sigma[a^s_i] \ \forall i \in I \Rightarrow  \sigma[t] \geq \sigma[s]
$$
where $I$ is the set of attributes of the tuples.\hfill \break
\begin{table}[h]
    \centering
    \begin{tabular}{| m{3cm} | m{3cm}| m{5cm} |  m{4cm} | } 
        \hline
        \textbf{Algorithm} & \textbf{Data access} & \textbf{Benefits} & \textbf{Drawbacks} \\ 
        \hline
        Fagin & sorted and random & easy to implement & cost is independent from the scoring function\\ 
        \hline
        Threshold & sorted and random & instance optimal & number of random access can be high \\
        \hline
        No Random Access & sorted & can be used when random access is unfeasible & scores might be uncertain \\
        \hline
       \end{tabular}
       \caption{top-k algorithm overview}
 \end{table}
\break
\textit{Fagin's algorithm} (\textit{FA}) \cite{FAGIN199983}: For each list of tuples make one sorted access at a time until \textit{k} objects in common are found.
For each selected object, using random access, apply the scoring function to compute its overall score, then select the best \textit{k} objects.\hfill \break
The complexity of the algorithm is \(O(N^{\frac{(m-1)}{m}}k^{\frac{1}{m}})\), where \textit{N} is the number of the tuples, \textit{m} 
is the number of the considered attributes and \textit{k} is the number of the best objects that we want to select.\hfill \break
An aspect that has to be kept into consideration, is that the access pattern of \textit{FA} is independent of the specified scoring function. Hence, the efficiency of the algorithm remains the same for all the possible scoring functions. 
Furthermore, since \textit{FA} must keep every object  of each sorted list in memory, its buffer is proportional to the size of the database.\hfill \break
\break
\textit{Threshold algorithm} (\textit{TA}) \cite{FAGIN2003614}: The \textit{TA} algorithm incrementally computes a threshold under which an object is excluded by the result set.
Unlike \textit{FA}, it performs parallel sorted access for each list, and for each extracted tuple,  it computes  its overall score through random access.
The threshold is updated at each iteration by applying the scoring function to the last objects seen in sorted access.
\textit{TA} terminates when the score of the \textit{k}-th object is above the threshold.\hfill \break
The number of sorted access performed by \textit{Threshold algorithm} is at most equal to that of \textit{FA}, but since it is instance optimal and thanks to the fact that it is strongly dependent on the shape of the scoring function, its performances are on average better than \textit{FA}.
In fact, \textit{TA} stops as soon as it has found the \textit{k}-best results.\hfill \break
Another advantage is that the algorithm uses a bounded buffer of size \textit{k}, which is independent of the size of the database.
Therefore, the amount of memory used by \textit{TA} is also optimal.
\hfill \break
\break
\textit{No Random Access algorithm} (\textit{NRA}) \cite{FAGIN2003614}: For each list of objects, makes one sorted access at a time and for each extracted tuple, keeps a sorted list with a threshold, a lower and an upper bound of the score. 
The algorithm stops when the \textit{k}-th object has a lower bound greater than the higher upper bound over all records.\hfill \break
Among all algorithms that don't perform random access, \textit{NRA} is instance optimal. 
Nonetheless, it needs an unlimited buffer to record the extracted objects with associated values. 
Another important aspect of the algorithm is that, for performance reasons, it provides only the \textit{k} best objects but unsorted, as information about the overall score is not presented to the user. 
Because random access is not available, looking for the exact order would require running the algorithm \textit{k} times. 
Assume that we want to find the best 3 objects, to do that we will execute \textit{NRA} 3 times, starting from \(k = 1\) and  increasing every time the value of \textit{k} by 1.
By doing this, we can keep a buffer of size \textit{k} where at each iteration we enqueue the object of the result set not yet present in the buffer.\hfill \break

\section{Skyline}
\label{sec:skyline}
\skyline offers a different approach to extract results from a dataset. 
Unlike \topk, this method provides to the user an overview of all optimal objects without  specifying a scoring function.
This framework is an extension of the \textit{maximum vector problem} \cite{Kung75onfinding, Godfrey2007} to the relational model, obtained by substituting vectors and points with a set of tuples.\hfill \break
The key concept behind \skyline is the notion of \textit{dominance}:\hfill \break
\break
\textit{Definition} Given two tuples \textit{t} and \textit{s} and a set of attributes \(A_i\) common to \textit{s} and \textit{t}, of cardinality \textit{m},
\textit{t} \textit{dominates} \textit{s} and indicate \(t \prec s\) if:
\[ s[A_i] \leq t[A_i] \quad \forall i \in \{1,..,m\} \] 
\[\exists \ j \ | \  s[A_j] < t[A_j] \quad  j \in \{1,..,m\} \] 
To explain it in another way, a tuple \textit{t} \textit{dominates} a tuple \textit{s} if all attributes of \textit{t} are at least equal to that of \textit{s} and if there exists at least one attribute \textit{t}
that is better than \textit{s}.\hfill \break
\skyline is defined as the set of tuples of a relation not dominated by any other tuples \cite{914855}.\hfill \break
As we said before,  to obtain the \skyline, we did not have to select any metric, because the result comes as a property of the dataset and for this reason, it presents to the users a broad view of the best objects. \hfill \break
On the other hand, the absence of a tool that expresses which attribute should be more important, it's usually not  the desired behavior, mainly with a large and fuzzy dataset. 
The size of the \textit{skyline} could be much bigger than the size of the result set obtained with a reasonable value of \textit{k} using the \topk method. \hfill \break
\break
\begin{figure}[h]
    \includegraphics[width=10cm]{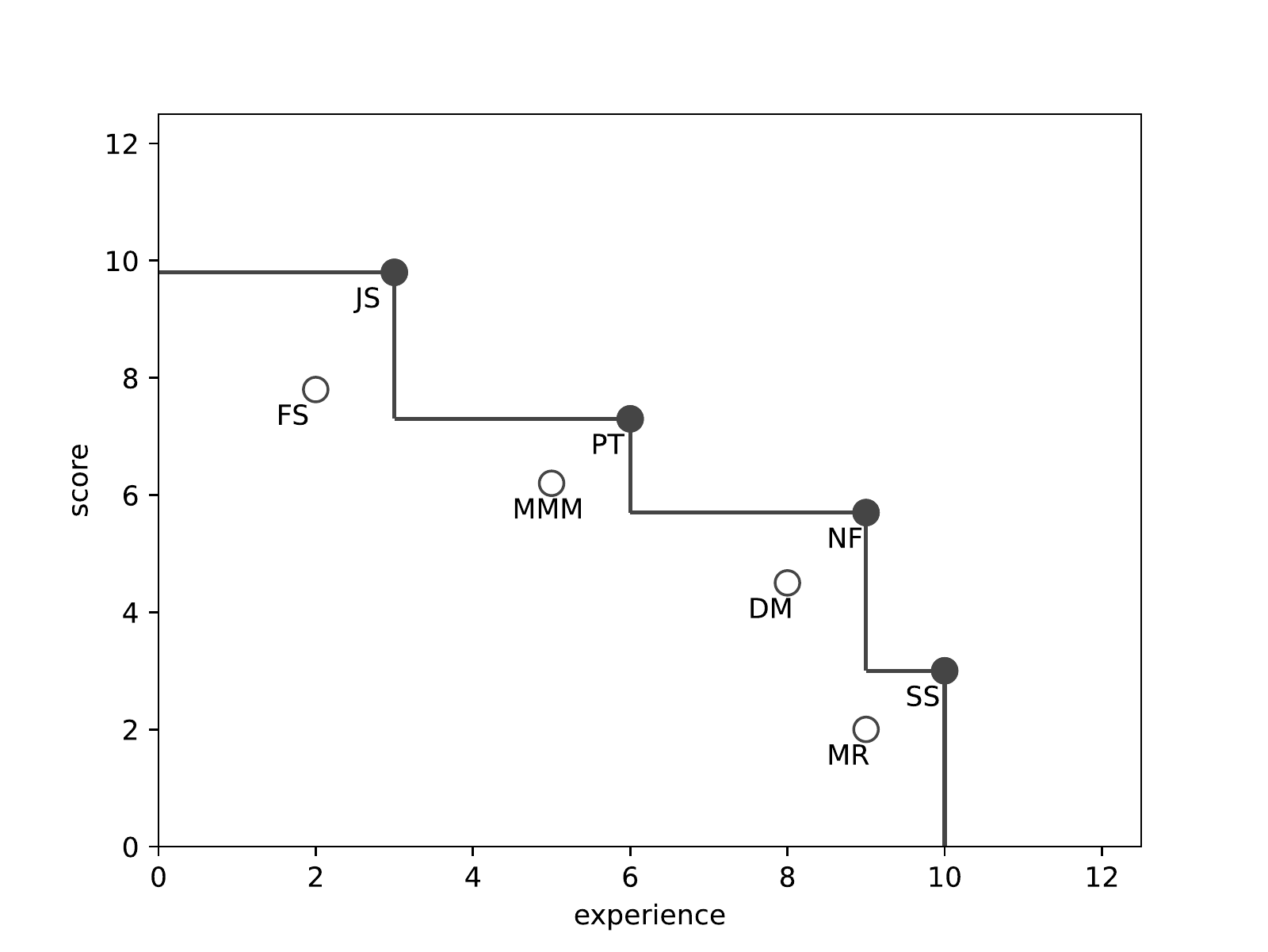}
    \centering
    \caption{\skyline of photographer dataset}
    \label{fig:exskyline}
\end{figure}
\break
\textit{Example} \hyperref[sec:ex1]{1} \textit{(cont.)}. 
Grey dots in Figure \hyperref[fig:exskyline]{4} show the skyline of the photographer dataset of Example \hyperref[sec:ex1]{1}.
Each object in the \textit{skyline} is not dominated by any other point; for example, JS dominates FS since both \textit{experience} and \textit{score} are greater than FS's ones (3 vs 2 and 9.8 vs 7.8).
Furthermore, JS is not dominated by PT, nor is PT  dominated by JS since from the \textit{experience} point of view, PT wins on JS but JS has a higher score with respect to PT. Thanks to the fact
that PT is not dominated by any other tuple, also PT is in the \textit{skyline}.\hfill \break

\subsection{Algorithms}

\textit{Block Nested Loop} (\textit{BNL}) \cite{914855}: the algorithm keeps track of all non-dominated points by inserting them in a buffer.
It scans all the dataset and compares each point \textit{p} with all points in the buffer: if \textit{p} is not dominated by any point, all points dominated by \textit{p} are removed from the buffer, and \textit{p} is added.\hfill \break
\textit{BNL} is as simple as inefficient since its effectiveness depends on the size of the buffer.
Considering the case, in which its size  is equal to the size of the dataset \textit{n}, the complexity of the algorithm is \(O(n^2)\).\hfill \break
\break
\textit{Soft Filter Skyline} (\textit{SFS}) \cite{1260846}: this algorithm is similar to the previous one, but \textit{SFS} starts by sorting the
dataset using a fixed monotone function (usually an entropy function). Due to the monotonicity of the sorting function, we know that given two sorted tuples \textit{t} and \textit{s}, if \textit{t} comes before
\textit{s}, \textit{t} is not \textit{dominated} by \textit{s}. As a consequence, if a tuple is in the buffer, it is a part of the skyline and it can be output without waiting for the end of the algorithm.\hfill \break
Furthermore, the number of iterations is optimal and the complexity is \(o(ln+n\log{n})\) where \textit{l} is the size of the skyline and
\textit{n} is the size of the dataset (\(O(n\log{n})\) needed for sorting).\hfill \break
In the worst case, the size of the \skyline equals the size of the dataset (\(l = n\)); the complexity of the \textit{SFS} turns back to \(O(n^2)\).\hfill \break

\section{Flexible-Skyline}
\label{sec:fskyline}
As already described in the upper sections, the two traditional  approaches have different pros and cons. Therefore, there is the need for a new method that can combine the best from \skyline and \topk and resolve the issues of both. \hfill \break 
\fskyline (\textit{F-skyline}) inherits from \topk the use of a scoring function but with an important difference: preferences are expressed in terms of constraints on the weights instead of letting the user guess the appropriate values. This solves one of the main weaknesses of \topk.\hfill \break
The major idea behind \fs, is that this framework outputs a subset of the \textit{skyline} of the dataset since it selects only the optimal objects that fulfill the constraints imposed by the scoring function.\hfill \break
This is done using the concept of \fdominance, which applies the \textit{dominance} to the \textit{scoring function}.\cite{10.1145/3406113, 10.14778/3137628.3137653} \hfill \break
\break
\textit{Definition} Given two distinct tuples \textit{t} and \textit{s}, and a family of monotone functions \(\mathcal{F}\), we said that
\textit{t} \textit{\(\mathcal{F}\)-dominates} \textit{s}, and indicate \(t \prec_\mathcal{F} s\), if
\[ f(s) \leq f(t) \quad \forall f \in \mathcal{F} \] 
In other words, \textit{t} \textit{\(\mathcal{F}\)-dominates} \textit{s} if there does not exist any scoring function in \(\mathcal{F}\) for which the value of \textit{s} is better than the value of \textit{t}.\hfill \break
\break
In particular, \fskyline offers two operators; given a set of tuples \textit{r},  \(\nd(r; \mathcal{F})\) is the subset of \textit{r} of tuples that are \textit{non}-$\mathcal{F}$-\textit{dominated} by any other tuples in \textit{r}, and \(\po(r; \mathcal{F})\) is the subset of \textit{r} of tuples that are potentially optimal according to a specific scoring function \(f \in \mathcal{F} \).
Considering that \nd and \po are both monotone operators with respect to the set of scoring functions, an important property is \cite{10.1145/3406113, 10.14778/3137628.3137653}:
\[ \po(r, \mathcal{F}) \subseteq \nd(r; \mathcal{F}) \subseteq \textsc{sky}(r)\] 
where \textsc{sky} is the \textit{skyline} of the dataset.\hfill \break 
This property highlights that \po can be obtained starting from \nd by looking for tuples that are not \textit{$\mathcal{F}$-dominated} by any convex combination of the tuples in \nd.
Therefore, \po outputs more accurate results than \nd and indicates that stricter constraints mean a smaller result set.
This is a significant result since it allows users to set the right levels of constraints according to their needs.\hfill \break
Another relevant concept is the notion of \fdominance \textit{region}, which is useful for computing both \po and \nd:\hfill \break
\break
\textit{Definition} The \fdominance \textit{region} of a tuple \textit{t} is the set of points that are  \textit{\(\mathcal{F}\)-dominated} by \textit{t}  \hfill \break
\break
\break
\begin{figure}[h]
    \centering
    \begin{minipage}{.5\textwidth}
      \centering
      \includegraphics[width=9cm]{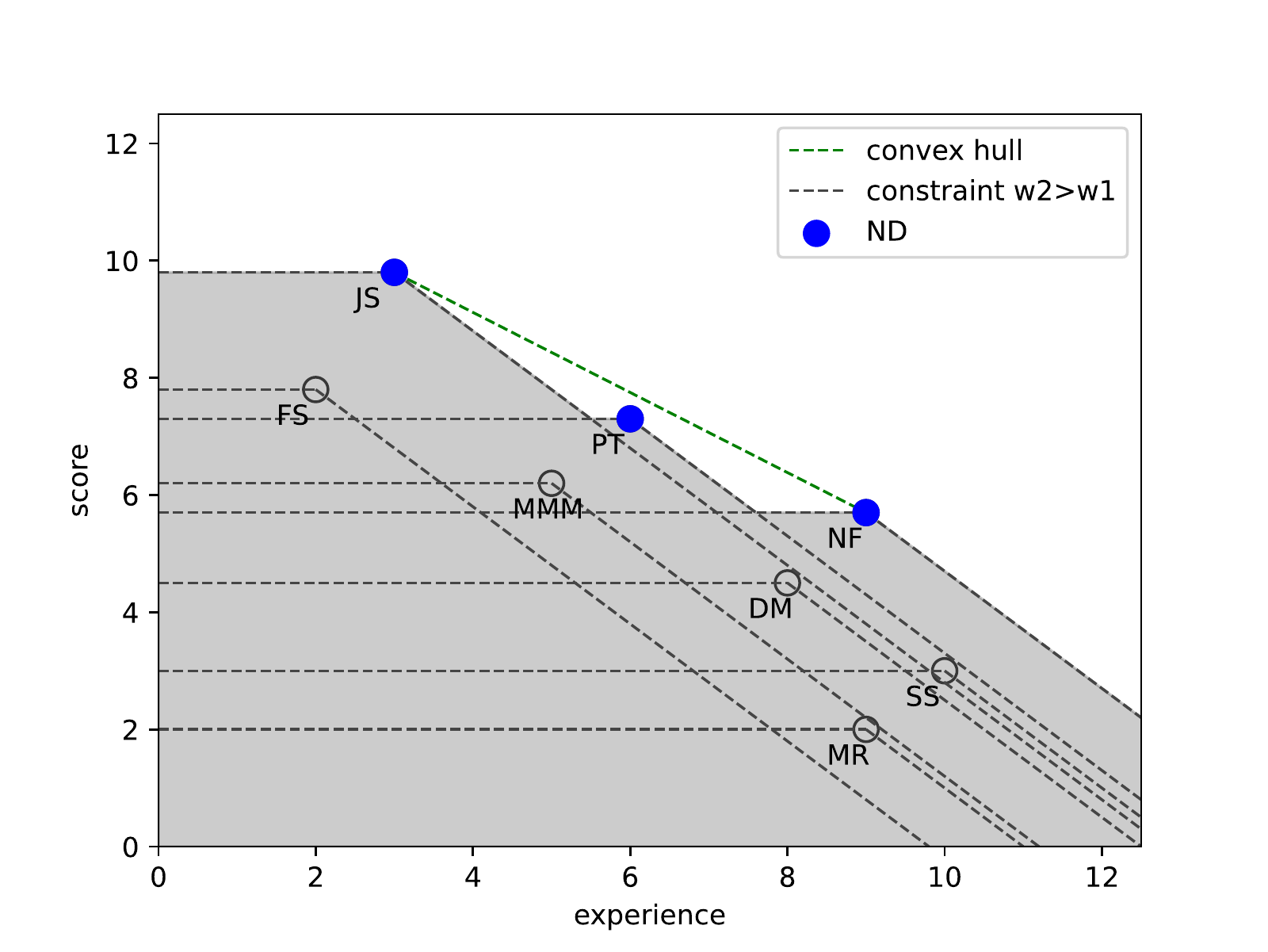}
      \caption{\nd set of \fs}
      \label{fig:exnd}
    \end{minipage}%
    \begin{minipage}{.5\textwidth}
      \centering
      \includegraphics[width=9cm]{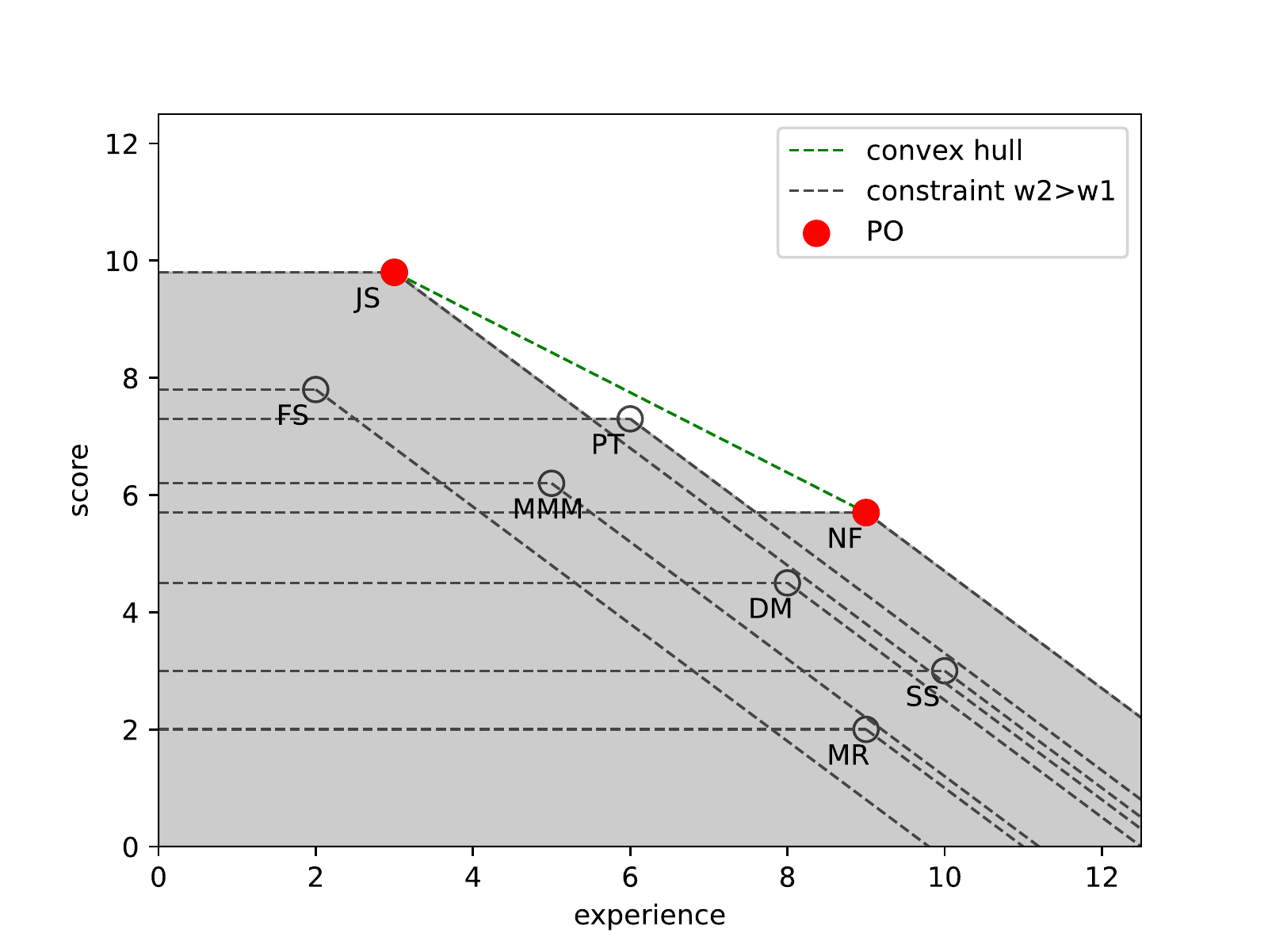}
      \caption{\po set of \fs}
      \label{fig:expo}
    \end{minipage}
\end{figure}
\break
\textit{Example} \hyperref[sec:ex1]{1} \textit{(cont.)}. Let us consider once again the situation shown before. The scoring function in its generic form is:
\[ F = w_1 · \ experience + w_2 · \ score \]
Assuming that the company prefers a photographer with a high score rather than experience, the choice of the weights using the \fs framework, will be in terms of constraints: \(w_2 > w_1\).\hfill \break
The grey area in Figure \hyperref[fig:exnd]{5} and Figure \hyperref[fig:expo]{6} represents  the unions of the \fdominance \textit{region} of all tuples, and as we expected, \nd is the subset of the \textit{skyline} that are not in such area. 
Furthermore, PT is not \textit{\(\mathcal{F}\)-dominated} by either JS and NF but it is below the hull obtained by the convex combination of the other tuples in \nd set.
That means that PT is \textit{\(\mathcal{F}\)-dominated} by all the virtual tuples on the hull and for this reason, it is not potentially optimal, so it is not in \po set.\hfill \break

\subsection{Algorithms}
For the implementation of \fs, several algorithms can be used to compute \po and \nd \cite{10.1145/3406113}, here we report the main characteristics and the most promising ones.\hfill \break
Concerning \nd, the classes of the algorithm are divided by:
\begin{itemize}
    \item \textit{Number of Phases}: \nd can be retrieved by computing \skyline (2 phases) or directly from the starting dataset (1 phase).
    \item \textit{Sorting}: the dataset can be sorted in advance according to a specific function (S) or not (U). Due to monotony, the advantage of sorting the dataset is that an upper tuple is never dominated by a lower one.
    \item \fdominance: \fdominance can be tested either by solving an associated \textit{LP problem} (LP) or by checking if the tuple is in \fdominance \textit{region} \cite{10.1145/3406113} using vertex enumeration (VE) of the polytope containing this region.
\end{itemize}
Concerning \po, the classes of the algorithm are divided by:
\begin{itemize}
    \item \textit{Phases}: \po can be retrieved by computing first \nd (2 phases) or directly from the dataset by excluding non-\po tuples (1 phase).
    \item \textit{\po test}: using primal \textit{\po test} (P) or the dual \po \textit{test} (D). The main difference is that the primal test exploits the dependence of \po to \nd, while the dual test uses the convex combination of a set of tuples. (See \cite{10.1145/3406113} for formal definitions)
    \item \textit{Incrementality}: each tuple can be tested with the set of all other tuples (U) or \textit{LP problem} of increasing sizes can be solved to test the tuple with a subset of the others (I). Using this last approach we can discard a tuple as soon as it is not optimal for a subset, due to the monotony of the functions \(f \in \mathcal{F} \).
\end{itemize}
Several experimental analyses \cite{10.1145/3406113} have verified that the best algorithm between the combination of the previous classes, is \textit{SVE1F} 
for computing \nd and \textit{PODI2} for computing \po.\hfill \break
\break
\textit{SVE1F} is a 1-phase algorithm that sorts the dataset before exploiting the vertex enumeration of the polytope representing the \fdominance \textit{region}. 
Its complexity is \(O(ve(c) + N · (\log{N} + |\nd| · q))\) where \(O(ve(c))\) is the cost for enumerating vertex (\textit{c} is the number of constraints), $|\nd|$ is the size of the set of the non-dominated tuples, \textit{q} is the number of vertices of the polytope and \textit{N} is the number of tuples in the dataset.\hfill \break
\break
\textit{PODI2} is a 2-phase algorithm that computes the \po set using the dual \po incrementally solving \textit{LP problems}.
Its complexity is \(C_{\nd} + O(|\nd| · \log{|\nd|} \ ·  lp(q, |\nd|))\). \(C_{\nd}\) is the complexity of the algorithm used for computing \nd set, that is, the starting point of the algorithm.
\(O(lp(q, |\nd|))\) is the complexity of computing the \textit{LP problem} with \textit{q} inequalities and \(|{\nd}|\) variables.\hfill \break

\subsection{Related work}
\fskyline was described for the first time under the name of \textit{Restricted-skyline} (\textit{R-skyline}) \cite{10.14778/3137628.3137653}, where only scoring functions in the form of $L_p$ norms were considered. \fs instead, extends the previous work letting the scoring function be subject to an arbitrary monotone transformation, the so called \textit{monotonically transformed, linear-in-the-weight} (MLW) functions.\hfill \break
\break
\textit{Uncertain top-k query} (\textit{UTK}) \cite{10.14778/3204028.3204031} is a method that offers a solution to the problem of weight selection in \topk. 
It aims is to report all possible \topk results to offer to the user similar options, starting from initial uncertain selected preferences. 
To do that, \textit{UTK} as \fs represents the user preferences in a convex polytope instead of a vector and inherits from \fskyline the concept of \fdominance. 
Furthermore, it extends \nd and \po operators applying them to \textit{k-skyband} \cite{10.1145/1061318.1061320} (generalization of the \skyline for $k \geq 1$).\hfill \break
\break
\textit{ORD} and \textit{ORU} operators \cite{10.1145/3448016.3457299} have the same goals as the previous ones but in addition, they can control the output size. This method starts from the preferences input vector of \topk, and expands it equally in all directions of the preferences domain.
As a consequence, additional records, that satisfy preferences similar to selected ones, are included in the result set. The \textit{OSS} (\textit{output-size specified}) property is accomplished by a stopping radius directly correlated to the desired output size.

\section{Conclusion}
\label{sec:comparison}
\topk is a very efficient way to extract data of interest from a dataset.
Unfortunately, the strength of this method is also its weakness; in fact, choosing the right weight values is a hard task,
that can lead to rerunning the queries several times by changing them. 
Furthermore, this method may not select some results near to the optimal ones that may be of interest to the user \cite{10.14778/3204028.3204031}.\hfill \break
On the other hand, using \skyline the user may have a general overview of the best results in the dataset, without selecting preferences on attributes.
This is paid in terms of the size of the result set: especially for a large anti-correlated dataset, the size of the \textit{skyline} can be too large that this method is unusable.\hfill \break
Furthermore, the simplicity of the algorithms for computing \skyline has the cost of a higher complexity: \(O(n^2)\) in the worst case.\hfill \break
\fskyline is a new approach for solving \textit{multi-objective} optimization problems that combine the properties of the previous methods.
It solves the problem of \topk by reducing user efforts in choosing the appropriate scoring function, meanwhile, it outputs a result set which size is a trade-off between the two previous approaches.
Moreover, it also offers two levels of accuracy thanks to \nd and \po operators. 
Despite such advantages, also \fs presents negative aspects. In particular, with this method, the user does not have control over the size of the output that could be larger or smaller than the user might expect \cite{10.1145/3448016.3457299}. 
Another limit of this operator is that the higher is the number of constraints, the more unstable is the \textit{LP solver}. In fact, for a large number of constraints, the result could be unreliable \cite{10.1145/3406113}.\hfill \break
Concerning complexity, \topk provides the most efficient algorithms among the others, even if \fs is comparable with \skyline.
Experimental results \cite{10.1145/3406113} show that \fskyline is more effective than \skyline, and for these reasons, it is a convenient alternative to those applications that require the characteristics of this approach.\hfill \break

\bibliographystyle{plain}
\bibliography{refs.bib}
\end{document}